\documentclass[aps,prl,twocolumn,superscriptaddress,showpacs]{revtex4}
\usepackage{float}
\usepackage{makeidx}
\usepackage{graphicx,amsmath}
\usepackage{CJK}
\usepackage{colordvi}
\usepackage{color}
\usepackage{amssymb}
\usepackage{hyperref}
\usepackage{graphicx}
\usepackage{dcolumn}
\usepackage{bm,amsmath,verbatim}
\usepackage{mathrsfs}
\usepackage{color}

\def\be{\begin{equation}}
\def\ee{\end{equation}}
\def\bea{\begin{eqnarray}}
\def\eea{\end{eqnarray}}
\def\bse{\begin{subequations}}
\def\ese{\end{subequations}}

\def\be{\begin{eqnarray}}
\def\ee{\end{eqnarray}}

\begin{document}

\title{Experimental observation of topological band gap opening in ultracold
Fermi gases with two-dimensional spin-orbit coupling }
\author{Zengming Meng}
\affiliation{State Key Laboratory of Quantum Optics and Quantum Optics Devices, Institute
of Opto-Electronics, Shanxi University, Taiyuan 030006, P.R.China }
\affiliation{Collaborative Innovation Center of Extreme Optics, Shanxi University,
Taiyuan 030006, P.R.China}
\author{Lianghui Huang}
\affiliation{State Key Laboratory of Quantum Optics and Quantum Optics Devices, Institute
of Opto-Electronics, Shanxi University, Taiyuan 030006, P.R.China }
\affiliation{Collaborative Innovation Center of Extreme Optics, Shanxi University,
Taiyuan 030006, P.R.China}
\author{Peng Peng}
\affiliation{State Key Laboratory of Quantum Optics and Quantum Optics Devices, Institute
of Opto-Electronics, Shanxi University, Taiyuan 030006, P.R.China }
\author{Donghao Li}
\affiliation{State Key Laboratory of Quantum Optics and Quantum Optics Devices, Institute
of Opto-Electronics, Shanxi University, Taiyuan 030006, P.R.China }
\author{Liangchao Chen}
\affiliation{State Key Laboratory of Quantum Optics and Quantum Optics Devices, Institute
of Opto-Electronics, Shanxi University, Taiyuan 030006, P.R.China }
\author{Yong Xu}
\affiliation{Department of Physics, The University of Texas at Dallas, Richardson, Texas
75080-3021, USA}
\author{Chuanwei Zhang}
\affiliation{Department of Physics, The University of Texas at Dallas, Richardson, Texas
75080-3021, USA}
\author{Pengjun Wang}
\affiliation{State Key Laboratory of Quantum Optics and Quantum Optics Devices, Institute
of Opto-Electronics, Shanxi University, Taiyuan 030006, P.R.China }
\affiliation{Collaborative Innovation Center of Extreme Optics, Shanxi University,
Taiyuan 030006, P.R.China}
\author{Jing Zhang}
\thanks{ Correspondence and requests for materials should be addressed to
J.Z. (email: jzhang74@sxu.edu.cn, jzhang74@yahoo.com).}
\affiliation{State Key Laboratory of Quantum Optics and Quantum Optics Devices, Institute
of Opto-Electronics, Shanxi University, Taiyuan 030006, P.R.China }
\affiliation{Synergetic Innovation Center of Quantum Information and Quantum Physics,
University of Science and Technology of China, Hefei, Anhui 230026, P. R.
China}

\begin{abstract}
The recent experimental realization of synthetic spin-orbit coupling (SOC)
opens a new avenue for exploring novel quantum states with ultracold atoms.
However, in experiments for generating two-dimensional SOC (e.g., Rashba
type), a perpendicular Zeeman field, which opens a band gap at the Dirac
point and induces many topological phenomena, is still lacking. Here we
theoretically propose and experimentally realize a simple scheme for
generating two-dimension SOC and a perpendicular Zeeman field simultaneously
in ultracold Fermi gases by tuning the polarization of three Raman lasers
that couple three hyperfine ground states of atoms. The resulting band gap
opening at the Dirac point is probed using spin injection radio-frequency
spectroscopy. Our observation may pave the way for exploring topological
transport and topological superfluids with exotic Majorana and Weyl fermion
excitations in ultracold atoms.
\end{abstract}

\maketitle

Spin-orbit coupling (SOC), the intrinsic interaction between a particle spin
and its motion, plays a key role in many important phenomena, ranging from
anomalous Hall effects \cite{Xiao} to topological insulators and
superconductors \cite{Hasan,Qi,Moore}. Although SOC is ubiquitous in nature,
the experimental control and observation of SOC induced effects are quite
difficult. In this context, the recent experimental realization of synthetic
SOC for cold atoms \cite%
{spielman,FuPRA,Shuai-PRL,Washington-PRA,Purdue,Jing,MIT,Spielman-Fermi}
provides a completely new and tunable platform for exploring SOC related
physics. Early experiments only realized the 1D SOC (\textit{i.e.}, an equal
sum of Rashba and Dresselhaus coupling, $\propto k_{x}\sigma _{y}$) using
two counter-propagating Raman lasers \cite%
{spielman,FuPRA,Shuai-PRL,Washington-PRA,Purdue,Jing,MIT,Spielman-Fermi}.
Many theoretical proposals have explored the generation of 2D SOC (\textit{%
i.e.}, $\propto \alpha k_{x}\sigma _{y}+\beta k_{y}\sigma _{x}$) \cite%
{Unanyan99PRA,Ruseckas:2005,Campbell11PRA,Xu13PRA,AndersonPRL,Yongping12PRL,Liu14PRL,Campbell15}
as well as their interesting physical properties in Bose and Fermi gases
\cite{WuCPL,Hui10PRL,Ho10PRL,Gong2011,Hui11PRL,Yu11PRL}. Recently, 2D SOC
was also experimentally realized in ultracold $^{40}$K Fermi gases \cite%
{Huang15} using three Raman lasers and the associated stable Dirac point on
a 2D momentum plane was observed \cite{Huang15}.

The experimental generation of SOC is usually accompanied with a Zeeman
field, which breaks various symmetries of the underlying system and induces
interesting quantum phenomena. The accompanied Zeeman field can be in-plane (%
\textit{e.g.}, $V\sigma _{y}$ for SOC $\propto k_{x}\sigma _{y}$) or
perpendicular (\textit{e.g.}, $V\sigma _{z}$ for SOC $\propto \alpha
k_{x}\sigma _{y}+\beta k_{y}\sigma _{x}$). The in-plane Zeeman field, while
preserves the Dirac point, makes the band dispersion asymmetric, leading to
new quantum states such as Fulde-Ferrell superfluids \cite%
{Zheng2013,Qu2013,Yi2013,Dong2013}. In contrast, the perpendicular Zeeman
field can open a topological band gap at the Dirac point of the SOC, leading
to many interesting topological transport \cite{Xiao} and superfluid
phenomena, such as the long-sought Majorana \cite{Zhang2008,Sau2010} and
Weyl \cite{Gong2011,Seo2012,Xu2014} fermions. In cold atom experiments,
although both in-plane and perpendicular Zeeman fields have been realized
with 1D SOC, only in-plane Zeeman field was realized with 2D SOC \cite%
{Huang15}. A perpendicular Zeeman field with 2D SOC is still lacked but
highly desired in experiments for the observation of various topological
transport and superfluid phenomena.

In this Letter, we theoretically propose and experimentally realize a simple
scheme for generating 2D SOC and a perpendicular Zeeman field
simultaneously. The same setup for generating 2D SOC in previous work \cite%
{Huang15} is used, in which three far-detuned Raman lasers couple three
hyperfine ground states. We only change the polarization of the Raman
lasers, which can create the perpendicular Zeeman field and open the band
gap at the Dirac point. This scheme significantly simplifies the
experimental requirement compared with existing theoretical proposals \cite%
{Zhang10PRA, Zhu11PRL}, which need additional laser beams, complicated
optical configuration, and controllable relative laser phases to create the
perpendicular Zeeman field. We observe and characterize the topological band
gap opening induced by the perpendicular Zeeman field using spin injection
radio-frequency (rf) spectroscopy, which are in good agreement with our
theoretical calculations.

\begin{figure}[tbp]
\includegraphics[width=8.5cm]{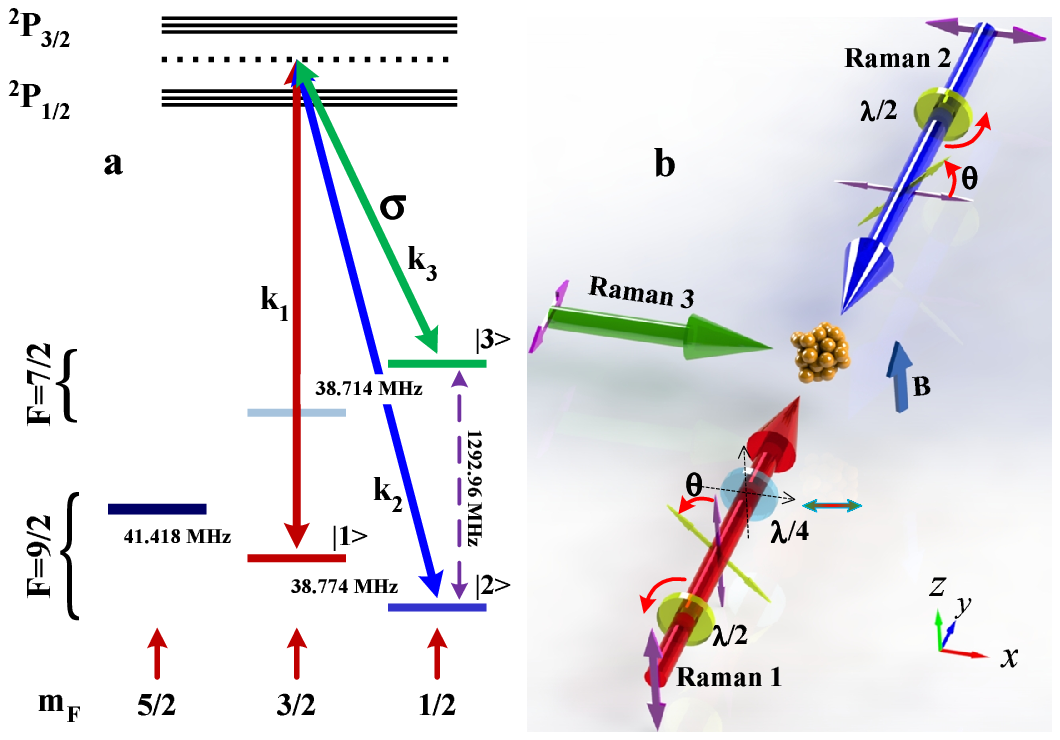}
\caption{Scheme of the atom-light interaction configuration for generating
2D synthetic SOC and an effective perpendicular Zeeman field simultaneously.
(\textbf{a}) Schematic of energy levels of $^{40}$K and Raman transitions.
Each of the three Raman lasers dresses one hyperfine ground state. (\textbf{b%
}) The experimental geometry and laser configuration. The Raman lasers 1-3
are initially prepared with the linear polarization along the \textit{z},
\textit{x} and \textit{y} directions, respectively. Then the Raman lasers 1
and 2 pass through a $\protect\lambda /2$ waveplate, which rotates their
linear polarizations by the same angle $\protect\theta $ synchronously.
Cases I and II correspond to without and with the $\protect\lambda /4$
waveplate after the $\protect\lambda /2$ waveplate for the Raman laser 1.}
\label{Fig1}
\end{figure}

\emph{Experimental setup and theoretical modelling:} The experimental setup
as well as theoretical modelling for generating and observing 2D SOC have
been described in details in our previous work \cite{Huang15}. In this
section we briefly describe our experimental system and corresponding
theoretical modelling with more details provided in the supplementary
information \cite{supp}. We point out that the lack of complex Raman
coupling strength is responsible for the lack of perpendicular Zeeman field
and the resulting topological band gap opening in our previous work \cite%
{Huang15}.

Specifically we consider ultracold $^{40}$K Fermi gases with three relevant
hyperfine states within the $4^{2}S_{1/2}$ ground electronic manifold, $%
|1\rangle =|F=9/2,m_{F}=3/2\rangle $, $|2\rangle =|F=9/2,m_{F}=1/2\rangle $,
and $|3\rangle =|F=7/2,m_{F}=1/2\rangle $, where ($F$, $m_{F}$) are the
quantum numbers for hyperfine ground states as shown in Fig. \ref{Fig1}a. A
homogeneous bias magnetic field $B_{0}=121.4$ G along the $z$ axis (the
gravity direction) produces a Zeeman shift to isolate these three hyperfine
states from others in the Raman transitions, as shown in Fig. \ref{Fig1}a.
We can neglect other hyperfine states and treat this system as one with
three ground states. Three far-detuned Raman lasers propagating on the $xy$
plane couple these three ground states to the electronically excited states
(Fig. \ref{Fig1}b).

For the far-detuned lasers, the excited states can be adiabatically
eliminated, and the Hamiltonian is written as $H=p_{z}^{2}/(2m)+H_{xy}$ with
\cite{Huang15}%
\begin{equation}
H_{xy}=\sum_{j=1}^{3}\left( \frac{(\mathbf{p}-\mathbf{k}_{j})^{2}}{2m}%
+\delta _{j}\right) |j\rangle \langle j|-\sum_{j^{\prime }\neq j}\frac{%
\Omega _{jj^{\prime }}}{2}|j\rangle \langle j^{\prime }|  \label{Hxy}
\end{equation}%
under the hyperfine ground state basis $\left\{ |j\rangle ,j=1,2,3\right\} $%
. Here $\mathbf{p}=p_{x}\mathbf{e}_{x}+p_{y}\mathbf{e}_{y}$ denotes the
momentum of atoms in the $xy$ plane, $\delta _{1}$ is set as zero (energy
reference) for simplification, $\delta _{2}$ ($\delta _{3}$) corresponds to
the two-photon Raman detuning between Raman lasers 1 and 2 (1 and 3). $%
\mathbf{k}_{1}=k_{r}\mathbf{e}_{y}$, $\mathbf{k}_{2}=-k_{r}\mathbf{e}_{y}$,
and $\mathbf{k}_{3}=k_{r}\mathbf{e}_{x}$ are the photon momenta of three
Raman lasers with the single-photon recoil momentum $k_{r}=2\pi \hbar
/\lambda $. Three $\Omega _{jj^{\prime }}=\Omega _{j^{\prime }j}^{\ast }$
describe the Raman coupling strength between hyperfine ground states $%
|j\rangle \leftrightarrow |j^{\prime }\rangle $ through Raman lasers. $k_{r}$
and recoil energy $E_{r}=k_{r}^{2}/2m$ are taken as natural momentum and
energy units.

The origin of 2D SOC in this system can be understood from the eigenstates
of the atom-laser interaction part (the non-diagonal terms $\Omega
_{jj^{\prime }}$) of the Hamiltonian (\ref{Hxy}), which contain two
degenerate dark states and one bright state separated by an energy gap \cite%
{supp}. Denote two degenerate dark states as the pseudo-spin states, the $%
3\times 3$ Hamiltonian (\ref{Hxy}) can be projected to the dark state
subspace, leading to an effective spin-half Hamiltonian with 2D SOC and
effective Zeeman fields. In previous experiments \cite{Huang15} as well as
existing theoretical proposals, $\Omega _{jj^{\prime }}$ are chosen to be
real numbers \cite{Note}, yielding
\begin{equation}
H_{eff}=\mathbf{p}^{2}/2m+H_{SOC}+V_{I},  \label{Heff}
\end{equation}%
where the 2D SOC $H_{SOC}=-\alpha p_{y}\sigma _{x}+\left( \beta
_{x}p_{x}-\beta _{y}p_{y}\right) \sigma _{y}$ and the in-plane Zeeman field $%
V_{I}=V_{x}\sigma _{x}+V_{y}\sigma _{y}$ in the $xy$ plane. $\alpha $, $%
\beta _{x}$, $\beta _{y}$, $V_{x}$, and $V_{y}$ are parameters determined by
the experimental parameters $\mathbf{k}_{j}\,$, $\delta _{j}$, and $\Omega
_{jj^{\prime }}$. Pauli matrices $\sigma _{i}$ are defined on the dark state
pseudo-spin basis. The in-plane Zeeman field $V_{I}$ shifts the Dirac cone
at $\mathbf{p=0}$ to another position in the momentum space, but cannot open
a band gap at the Dirac point. Such controlled shift of the Dirac point has
been observed in our experiment \cite{Huang15}. However, the opening of a
topological band gap at the Dirac point requires a perpendicular Zeeman
field of the form $V_{z}\sigma _{z}$, which demands the generation of
complex Raman coupling $\Omega _{jj^{\prime }}$ in experiments.

\emph{Realization of a perpendicular Zeeman field:} Because the Raman
coupling strengths $\Omega _{jj^{\prime }}$ are proportional to$\
\overrightarrow{E}_{j}\times \overrightarrow{E}_{j^{\prime }}$, the
generation of complex $\Omega _{jj^{\prime }}$ requires the tuning of the
polarizations of Raman lasers from linear to elliptical, which can be
realized in experiments using $\lambda /2$ and $\lambda /4$ waveplates.

The scheme for inserting $\lambda /2$ and $\lambda /4$ waveplates in the experimental
setup is shown in Fig. \ref{Fig1}(b). Here two $\lambda /2$ waveplates
rotate the polarizations of Raman lasers 1 and 2 by an angle $\theta $,
yielding $\overrightarrow{E}_{1}=A_{1}(\cos \theta \hat{\mathbf{e}}%
_{\parallel }+\sin \theta \hat{\mathbf{e}}_{\bot })$ (without the $\lambda /4
$ waveplate) and $\overrightarrow{E}_{2}=A_{2}(\cos \theta \hat{\mathbf{e}}%
_{\bot }+\sin \theta \hat{\mathbf{e}}_{\parallel })$, where $\hat{\mathbf{e}}%
_{\bot }$ and $\hat{\mathbf{e}}_{\parallel }$ components correspond to $%
\sigma $ and $\pi $ polarizations with respect to the quantization axis
\textbf{z} defined by the magnetic field. The rotation still keeps $%
\overrightarrow{E}_{1}$ and $\overrightarrow{E}_{2}$ orthogonal. The
additional $\lambda /4$ waveplate can change $\overrightarrow{E}_{1}$ to
elliptical polarization $\overrightarrow{E}_{1}=A_{1}(\cos \theta \hat{%
\mathbf{e}}_{\parallel }+i\sin \theta \hat{\mathbf{e}}_{\perp })$, where the
imaginary part is responsible for generating complex $\Omega _{12}$. To
illustrate this, we consider two different cases (I) without and (II) with
the $\lambda /4$ waveplate. Hereafter we denote $\Omega _{jj^{\prime }}$ as
the Raman coupling strength before the $\lambda /2$ and $\lambda /4$
waveplates. Careful analysis of the Raman transition selection rules (note
that $\overrightarrow{E}_{3}=A_{3}\hat{\mathbf{e}}_{\bot }$) shows \cite%
{supp}:

\begin{figure}[t]
\includegraphics[width=8.5cm]{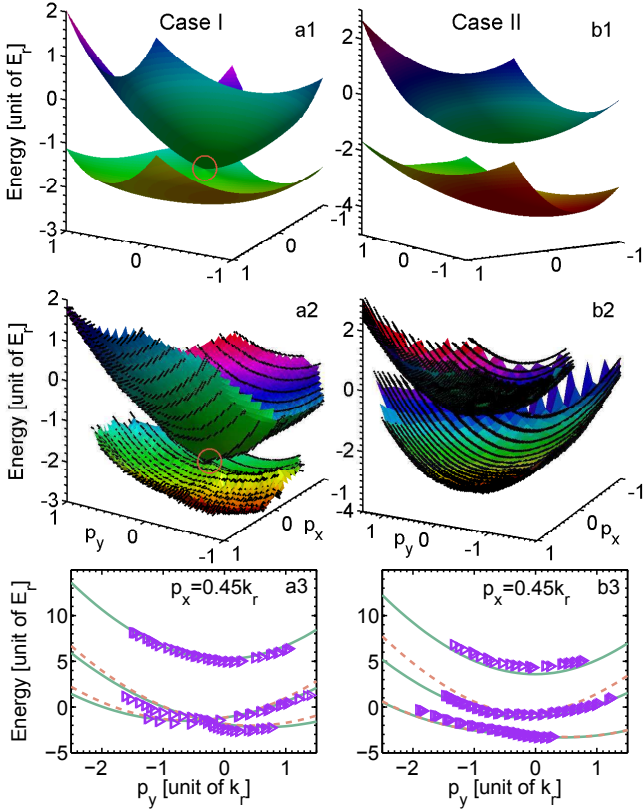}
\caption{The energy dispersions of dressed atoms measured by rf
spin-injection spectroscopy. Columns \textbf{(a1)-(a3)} and (\textbf{b1)}-(%
\textbf{b3)} correspond to the energy-momentum dispersions of 2D SOC without
(case I) and with (case II) effective perpendicular Zeeman field,
respectively. The experimental parameters are $\Omega _{12}=-4.97E_{r}$, $%
\Omega _{13}=5.46E_{r}$, $\Omega _{23}=6.46E_{r}$, $\protect\delta %
_{2}=-0.5E_{r}$, $\protect\delta _{3}=-1.8E_{r}$ and $\protect\theta =45^{o}$%
. (\textbf{a1)} and (\textbf{b1)} are theoretical results calculated using
the Hamiltonian (\protect\ref{Hxy}) with the experimental parameters. (%
\textbf{a2)} and (\textbf{b2)} are experimental results measured by rf
spin-injection spectroscopy. The black dots represent the experimental data.
The yellow circles in (\textbf{a1}) and (\textbf{a2}) indicate the Dirac
points. In both rows, we only show the lowest two bands for better
visualization of the Dirac points and the band gap opening. (\textbf{a3)}
The cross-section drawings of (\textbf{a1}) and (\textbf{a2}) in the energy-$%
p_{y}$ coordinates for $p_{x}=0.45k_{r}$. Triangles are from experimental
data. Solid and dashed lines are from theoretical calculations using the
full (Eq. \protect\ref{Hxy}) and effective (Eq. \protect\ref{Heff})
Hamiltonians, respectively. (\textbf{b3)}, The cross-section drawings of
(b1) and (b2). }
\label{Fig2}
\end{figure}

\textbf{Case I}: $\Omega _{23}^{I}=\cos \theta \Omega _{23}$, $\Omega
_{13}^{I}=\cos \theta \Omega _{13}$, and $\Omega _{12}^{I}=\Omega _{12}$.
The rotations induced by the $\lambda /2$ waveplates keep $\Omega
_{jj^{\prime }}^{I}$ real, therefore only shift the Dirac point position and
cannot open a band gap (see Fig. \ref{Fig2}.a1).

\textbf{Case II:} $\Omega _{13}^{II}=\cos \theta \Omega _{13}$, $\Omega
_{23}^{II}=\cos \theta \Omega _{23}$, and $\Omega _{12}^{II}=\Omega
_{12}(\cos ^{2}\theta +i\sin ^{2}\theta )$, yielding an imaginary part $%
H_{Z}=-i\frac{\Omega _{12}\sin ^{2}\theta }{2}|1\rangle \langle 2|+H.c.$ in
the Hamiltonian (\ref{Hxy}), which cannot be gauged out by varying the phase
of the wavefunction for each hyperfine ground state. This term opens a band
gap at the Dirac point as shown in Fig. \ref{Fig2}.b1. In the degenerate
dark state pseudo-spin basis, this term gives $V_{z}\sigma _{z}$, the
perpendicular Zeeman field. The energy gap at the Dirac point can be
controlled precisely by adjusting the rotation angle $\theta $. Note $H_{Z}$
has the same form as that in previous theoretical proposal \cite{Zhu11PRL}
that requires complicated setup of additional lasers. Our scheme is much
simpler and more robust because it only need tune the polarizations of three
Raman lasers.

\begin{figure}[t]
\includegraphics[width=8.5cm]{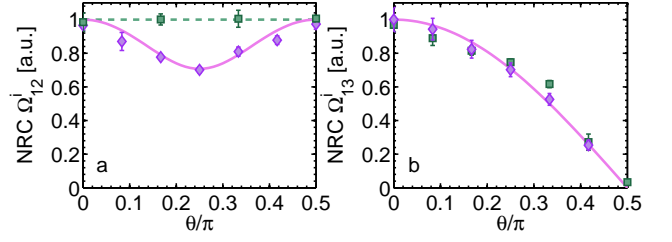}
\caption{Measure the Raman coupling strength by Rabi oscillations between
two hyperfine ground states. Plot of the Raman coupling strengths $\Omega
_{12}^{i}$ (a) and $\Omega _{13}^{i}$ (b) versus $\protect\theta $. The
green squares and purple diamonds correspond to cases I and II respectively.
NRC represents normalized Raman coupling. The solid and dashed lines are
theoretical curves. }
\label{Fig3}
\end{figure}

The change of $\Omega _{jj^{\prime }}^{i}$ induced by the waveplates can be
measured using the Rabi oscillation between two hyperfine ground states \cite%
{Jing,supp}. We obtain $\Omega _{12}^{i}$ for cases I and II respectively as
the function of the rotation angle $\theta $ as shown in Fig. \ref{Fig3}a.
For case I, $\Omega _{12}^{I}$ keeps unchanged for different $\theta $. In
case II, we measure the absolute value of $\Omega _{12}^{II}$ because $%
\Omega _{12}^{II}$ is a complex number, which shows $\sqrt{\cos ^{4}\theta
+\sin ^{4}\theta }$ dependence for different $\theta $ \ (Fig. \ref{Fig3}a),
agreeing with the theory. The measurements for $\Omega _{13}^{i}$ and $%
\Omega _{23}^{i}$ also demonstrate their $\cos \theta $ dependence (Fig. \ref%
{Fig3}b).

\emph{Observation of band gap opening:} The imaginary part $i\Omega
_{12}\sin ^{2}\theta $ of $\Omega _{12}^{II}$ opens a band gap at the Dirac
point of the energy-momentum dispersions, corresponding to a perpendicular
Zeeman field. Such topological band gap opening can be measured by spin
injection rf spectroscopy, which uses rf field to drive atoms from a free
spin-polarized state into an empty 2D SOC system \cite{Huang15}.

In our experiment, a degenerate Fermi gas ${}^{40}$K of $2\times 10^{6}$ is
prepared at the free reservoir hyperfine state $|9/2,5/2\rangle $ in a
crossed optical dipole trap. We ramp the homogeneous bias magnetic field to
the value $B_{0}=121.4$ G, and then apply three Raman lasers with the
wavelength 768.85 nm in 60 ms from zero to its final value. Subsequently, a
Gaussian shape pulse of the rf field is applied for 450 $\mu s$ to drive
atoms from the initial $|9/2,5/2\rangle $ state to the final empty state
with 2D SOC. Since rf field does not transfer momentum to the atoms, spin
injection occurs when the frequency of the rf matches the energy difference
between the initial and final states \cite{supp}. At last, the Raman lasers,
the optical trap and the bias magnetic field are switched off abruptly, and
atoms freely expand for 12 ms with a magnetic field gradient applied along
the \textit{x} axis. Absorption image is taken along the $\mathit{z}$
direction. We use a Gaussian fit to locate the maximum of the measured
atomic density as a function of the momentum and the rf frequency from the
absorption image, from which we can obtain the energy band dispersion.

Fig. \ref{Fig2} shows the momentum-resolved spin-injection spectra for cases
I and II. Other experimental parameters are the same for two cases. When the
wavelength of the Raman lasers is tuned to 768.85 nm between the $D_{1}$ and
$D_{2}$ lines, two lower energy dispersions touch at a Dirac point for case
I as shown in Figs. \ref{Fig2}(a2)-(a3), which demonstrate the 2D SOC \cite%
{Huang15}. With the $\lambda /4$ waveplate (case II), the energy gap at the
Dirac point is opened, as shown in Fig. \ref{Fig2}(b2)-(b3). We perform
numerical calculations for the energy spectra of the Hamiltonian (\ref{Hxy})
and the effective Hamiltonian (\ref{Heff}) with corresponding experimental
parameters, which show good agreement with the experimental data. Note that
the eigenenergy of the effective Hamiltonian (\ref{Heff}) deviates from the
exact Hamiltonian (\ref{Hxy}) for momentum away from the Dirac points.

\begin{figure}[t]
\includegraphics[width=3.4in]{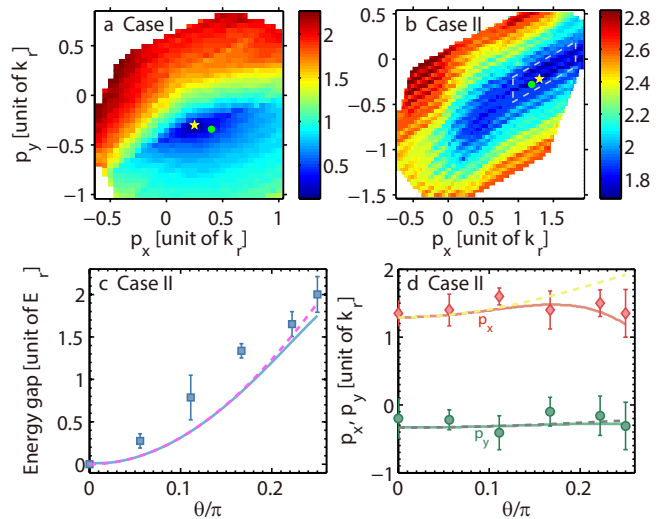}
\caption{Tunable band gap at the Dirac point induced by the perpendicular
Zeeman field. (\textbf{a,b}) Plot of the energy differences between the
lowest two bands in experiments. The stars and dots correspond to the Dirac
point positions in experiments and theory, respectively. (\textbf{c, d}) The
band gap (\textbf{c}) and the position of the Dirac point (\textbf{d}) are
plotted as the function of $\protect\theta $. In both figures, the squares
and diamonds represent experimental data. The solid and dashed lines
represent theoretical calculations using the $3\times 3$ Hamiltonian (%
\protect\ref{Hxy}) and the spin-half effective Hamiltonian (\protect\ref%
{Heff}) respectively. The experimental parameters are the same as Fig.
\protect\ref{Fig2}. }
\label{Fig4}
\end{figure}

In experiments, we determine the band gap and the position of the Dirac
point by searching the minimum of the energy differences between the lowest
two bands in experiments, as shown in Fig. \ref{Fig4}(a,b) for two cases
with $\theta =\pi /4$. The stars and dots represent the positions of band
gap minima measured in experiments and obtained in theory using the exact $%
3\times 3$ Hamiltonian (\ref{Hxy}) \cite{supp}, respectively. The band gap
is very large for a large $\theta =\pi /4$, therefore the bands are very
flat around the gap minima, yielding large uncertainty for the measurement
of Dirac point positions in experiments (see the dashed line box in Fig. \ref%
{Fig4}b).

Fig. \ref{Fig4}(c,d) shows the band gap and Dirac point positions as a
function of $\theta $. We see the band gap increases with the angle $\theta $%
, while the positions of the Dirac points only change slightly,
demonstrating the tunability of the topological band gaps through varying
the polarization of the Raman laser. The measured single particle band gaps
and Dirac point position do not exactly agree with theoretical calculations
from the Hamiltonians (\ref{Hxy}) and (\ref{Heff}), which may be attributed
to, for instance, the finite energy resolution of rf spectrum and the
corresponding finite momentum width, the uncertainty in the Gaussian fit
process to locate the maximum of the measured atomic density that determines
the atom momentum, and the stability of the magnetic field, \textit{etc}.

\emph{Discussion}: The topological properties of the induced band gap by the
perpendicular Zeeman field can be characterized by the Berry curvature of
each band \cite{Xiao} (see supplementary information for the plot of the
Berry curvatures in the lowest two bands), which is a delta function at the
Dirac point for case I, but becomes nonzero for all $\mathbf{p}$ with a peak
located at the Dirac point for case II. The corresponding Berry phases are
found to be $\mp \pi $ for the lowest two bands, as expected.

By varying the Raman laser intensities in experiments, the positions of the
Dirac points, the form of the 2D SOC, and the associated in-plane Zeeman
field can be tuned. Together with the tunable perpendicular Zeeman field,
our system provides a potential platform for exploring various SOC related
transport phenomena for non-interacting atoms and superfluid physics for
Cooper pairs. For instance, the coexistence of 2D SOC and perpendicular
Zeeman field yields non-zero Berry curvature, leading to an anomalous
velocity for atoms \cite{Xiao}. The resulting anomalous Hall effects may be
observed in a non-interacting Fermi gas \cite{Zhang10PRA}. The generated
perpendicular Zeeman field is at the order of recoil energy $E_{r}$, which
is large enough for realizing topological superfluids \cite%
{Zhang2008,Sau2010} and associated Majorana and Weyl fermions \cite%
{Gong2011,Seo2012,Xu2014}. The generated Zeeman field contains both in-plane
and perpendicular components, which make the single particle band structure
highly asymmetric, leading to the possibility of observing the long-sought
Fulde-Ferrell-Larkin-Ovchinnikov superfluid phases with finite momentum
pairing \cite{Zheng2013,Qu2013,Yi2013}. In particular, in the two-body
physics level, such system could host a dimer bound state with finite
center-of-mass mechanical momentum, which may be measured in the momentum
distribution \cite{Dong2013}.

In summary, we have realized a simple scheme for generating 2D SOC and a
perpendicular Zeeman field simultaneously for ultracold fermionic atoms. The
topological energy gap at the Dirac point can be opened and controlled
precisely by the perpendicular Zeeman field. Our study should pave the way
for exploring various interesting topological and other exotic superfluid
phenomena arising from the \textit{s}-wave scattering interaction.

\begin{acknowledgments}
This research is supported by the NSFC (Grant No. 11234008, 11361161002,
11222430) and the program for Sanjin Scholars of Shanxi Province. Y.X. and
C.Z. are supported by ARO (W911NF-12-1-0334), AFOSR (FA9550-16-1-0387), and
NSF (PHY-1505496).
\end{acknowledgments}


\begin{thebibliography}{99}
\bibitem{Note} Note that although different constant phases can be
introduced for each Raman laser, which make all $\Omega _{jj^{\prime }}$
complex, such phases can be easily gauged out by absorbing the phases into
the definition of each hyperfine state wavefunction, yielding real $\Omega
_{jj^{\prime }}$.
\end{thebibliography}
\end{document}